\documentclass[aps,prd,twocolumn,showpacs,amsmath,amssymb,reprint,superscriptaddress]{revtex4}

\usepackage{times}
\usepackage{mathptm} 
\usepackage{fancyhdr}
\usepackage{mdwlist}
\usepackage{graphicx}
\usepackage{epstopdf}
\usepackage{epsfig}
\usepackage{wrapfig}
\usepackage{multirow} 

\usepackage{graphicx}
\usepackage{dcolumn}
\usepackage{bm}

\usepackage{helvet}

\begin{document}


\title{Exclusion of exotic top-like quarks with --4/3 electric charge using jet-charge tagging in single-lepton $t\bar{t}$ events at CDF}

\affiliation{Institute of Physics, Academia Sinica, Taipei, Taiwan 11529, Republic of China}
\affiliation{Argonne National Laboratory, Argonne, Illinois 60439, USA}
\affiliation{University of Athens, 157 71 Athens, Greece}
\affiliation{Institut de Fisica d'Altes Energies, ICREA, Universitat Autonoma de Barcelona, E-08193, Bellaterra (Barcelona), Spain}
\affiliation{Baylor University, Waco, Texas 76798, USA}
\affiliation{Istituto Nazionale di Fisica Nucleare Bologna, $^{ee}$University of Bologna, I-40127 Bologna, Italy}
\affiliation{University of California, Davis, Davis, California 95616, USA}
\affiliation{University of California, Los Angeles, Los Angeles, California 90024, USA}
\affiliation{Instituto de Fisica de Cantabria, CSIC-University of Cantabria, 39005 Santander, Spain}
\affiliation{Carnegie Mellon University, Pittsburgh, Pennsylvania 15213, USA}
\affiliation{Enrico Fermi Institute, University of Chicago, Chicago, Illinois 60637, USA}
\affiliation{Comenius University, 842 48 Bratislava, Slovakia; Institute of Experimental Physics, 040 01 Kosice, Slovakia}
\affiliation{Joint Institute for Nuclear Research, RU-141980 Dubna, Russia}
\affiliation{Duke University, Durham, North Carolina 27708, USA}
\affiliation{Fermi National Accelerator Laboratory, Batavia, Illinois 60510, USA}
\affiliation{University of Florida, Gainesville, Florida 32611, USA}
\affiliation{Laboratori Nazionali di Frascati, Istituto Nazionale di Fisica Nucleare, I-00044 Frascati, Italy}
\affiliation{University of Geneva, CH-1211 Geneva 4, Switzerland}
\affiliation{Glasgow University, Glasgow G12 8QQ, United Kingdom}
\affiliation{Harvard University, Cambridge, Massachusetts 02138, USA}
\affiliation{Division of High Energy Physics, Department of Physics, University of Helsinki and Helsinki Institute of Physics, FIN-00014, Helsinki, Finland}
\affiliation{University of Illinois, Urbana, Illinois 61801, USA}
\affiliation{The Johns Hopkins University, Baltimore, Maryland 21218, USA}
\affiliation{Institut f\"{u}r Experimentelle Kernphysik, Karlsruhe Institute of Technology, D-76131 Karlsruhe, Germany}
\affiliation{Center for High Energy Physics: Kyungpook National University, Daegu 702-701, Korea; Seoul National University, Seoul 151-742, Korea; Sungkyunkwan University, Suwon 440-746, Korea; Korea Institute of Science and Technology Information, Daejeon 305-806, Korea; Chonnam National University, Gwangju 500-757, Korea; Chonbuk National University, Jeonju 561-756, Korea; Ewha Womans University, Seoul, 120-750, Korea}
\affiliation{Ernest Orlando Lawrence Berkeley National Laboratory, Berkeley, California 94720, USA}
\affiliation{University of Liverpool, Liverpool L69 7ZE, United Kingdom}
\affiliation{University College London, London WC1E 6BT, United Kingdom}
\affiliation{Royal Holloway, University of London, Egham, Surrey, TW20 0EX, UK}
\affiliation{Centro de Investigaciones Energeticas Medioambientales y Tecnologicas, E-28040 Madrid, Spain}
\affiliation{Massachusetts Institute of Technology, Cambridge, Massachusetts 02139, USA}
\affiliation{Institute of Particle Physics: McGill University, Montr\'{e}al, Qu\'{e}bec H3A~2T8, Canada; Simon Fraser University, Burnaby, British Columbia V5A~1S6, Canada; University of Toronto, Toronto, Ontario M5S~1A7, Canada; and TRIUMF, Vancouver, British Columbia V6T~2A3, Canada}
\affiliation{University of Michigan, Ann Arbor, Michigan 48109, USA}
\affiliation{Michigan State University, East Lansing, Michigan 48824, USA}
\affiliation{Institution for Theoretical and Experimental Physics, ITEP, Moscow 117259, Russia}
\affiliation{University of New Mexico, Albuquerque, New Mexico 87131, USA}
\affiliation{The Ohio State University, Columbus, Ohio 43210, USA}
\affiliation{Okayama University, Okayama 700-8530, Japan}
\affiliation{Osaka City University, Osaka 588, Japan}
\affiliation{University of Oxford, Oxford OX1 3RH, United Kingdom}
\affiliation{Istituto Nazionale di Fisica Nucleare, Sezione di Padova-Trento, $^{ff}$University of Padova, I-35131 Padova, Italy}
\affiliation{University of Pennsylvania, Philadelphia, Pennsylvania 19104, USA}
\affiliation{Istituto Nazionale di Fisica Nucleare Pisa, $^{gg}$University of Pisa, $^{hh}$University of Siena and $^{ii}$Scuola Normale Superiore, I-56127 Pisa, Italy, $^{mm}$INFN Pavia and University of Pavia, I-27100 Pavia, Italy}
\affiliation{University of Pittsburgh, Pittsburgh, Pennsylvania 15260, USA}
\affiliation{Purdue University, West Lafayette, Indiana 47907, USA}
\affiliation{University of Rochester, Rochester, New York 14627, USA}
\affiliation{The Rockefeller University, New York, New York 10065, USA}
\affiliation{Istituto Nazionale di Fisica Nucleare, Sezione di Roma 1, $^{jj}$Sapienza Universit\`{a} di Roma, I-00185 Roma, Italy}
\affiliation{Mitchell Institute for Fundamental Physics and Astronomy, Texas A\&M University, College Station, Texas 77843, USA}
\affiliation{Istituto Nazionale di Fisica Nucleare Trieste/Udine; $^{nn}$University of Trieste, I-34127 Trieste, Italy; $^{kk}$University of Udine, I-33100 Udine, Italy}
\affiliation{University of Tsukuba, Tsukuba, Ibaraki 305, Japan}
\affiliation{Tufts University, Medford, Massachusetts 02155, USA}
\affiliation{University of Virginia, Charlottesville, Virginia 22906, USA}
\affiliation{Waseda University, Tokyo 169, Japan}
\affiliation{Wayne State University, Detroit, Michigan 48201, USA}
\affiliation{University of Wisconsin, Madison, Wisconsin 53706, USA}
\affiliation{Yale University, New Haven, Connecticut 06520, USA}

\author{T.~Aaltonen}
\affiliation{Division of High Energy Physics, Department of Physics, University of Helsinki and Helsinki Institute of Physics, FIN-00014, Helsinki, Finland}
\author{S.~Amerio}
\affiliation{Istituto Nazionale di Fisica Nucleare, Sezione di Padova-Trento, $^{ff}$University of Padova, I-35131 Padova, Italy}
\author{D.~Amidei}
\affiliation{University of Michigan, Ann Arbor, Michigan 48109, USA}
\author{A.~Anastassov$^x$}
\affiliation{Fermi National Accelerator Laboratory, Batavia, Illinois 60510, USA}
\author{A.~Annovi}
\affiliation{Laboratori Nazionali di Frascati, Istituto Nazionale di Fisica Nucleare, I-00044 Frascati, Italy}
\author{J.~Antos}
\affiliation{Comenius University, 842 48 Bratislava, Slovakia; Institute of Experimental Physics, 040 01 Kosice, Slovakia}
\author{G.~Apollinari}
\affiliation{Fermi National Accelerator Laboratory, Batavia, Illinois 60510, USA}
\author{J.A.~Appel}
\affiliation{Fermi National Accelerator Laboratory, Batavia, Illinois 60510, USA}
\author{T.~Arisawa}
\affiliation{Waseda University, Tokyo 169, Japan}
\author{A.~Artikov}
\affiliation{Joint Institute for Nuclear Research, RU-141980 Dubna, Russia}
% ...

\collaboration{CDF Collaboration\footnote{With visitors from
$^a$University of British Columbia, Vancouver, BC V6T 1Z1, Canada,
$^b$Istituto Nazionale di Fisica Nucleare, Sezione di Cagliari, 09042 Monserrato (Cagliari), Italy,
$^c$University of California Irvine, Irvine, CA 92697, USA,
$^e$Institute of Physics, Academy of Sciences of the Czech Republic, 182~21, Czech Republic,
$^f$CERN, CH-1211 Geneva, Switzerland,
$^g$Cornell University, Ithaca, NY 14853, USA,
$^{dd}$The University of Jordan, Amman 11942, Jordan,
$^h$University of Cyprus, Nicosia CY-1678, Cyprus,
$^i$Office of Science, U.S. Department of Energy, Washington, DC 20585, USA,
$^j$University College Dublin, Dublin 4, Ireland,
$^k$ETH, 8092 Z\"{u}rich, Switzerland,
$^l$University of Fukui, Fukui City, Fukui Prefecture, Japan 910-0017,
$^m$Universidad Iberoamericana, Lomas de Santa Fe, M\'{e}xico, C.P. 01219, Distrito Federal,
$^n$University of Iowa, Iowa City, IA 52242, USA,
$^o$Kinki University, Higashi-Osaka City, Japan 577-8502,
$^p$Kansas State University, Manhattan, KS 66506, USA,
$^q$Brookhaven National Laboratory, Upton, NY 11973, USA,
$^r$University of Manchester, Manchester M13 9PL, United Kingdom,
$^s$Queen Mary, University of London, London, E1 4NS, United Kingdom,
$^t$University of Melbourne, Victoria 3010, Australia,
$^u$Muons, Inc., Batavia, IL 60510, USA,
$^v$Nagasaki Institute of Applied Science, Nagasaki 851-0193, Japan,
$^w$National Research Nuclear University, Moscow 115409, Russia,
$^x$Northwestern University, Evanston, IL 60208, USA,
$^y$University of Notre Dame, Notre Dame, IN 46556, USA,
$^z$Universidad de Oviedo, E-33007 Oviedo, Spain,
$^{aa}$CNRS-IN2P3, Paris, F-75205 France,
$^{cc}$Universidad Tecnica Federico Santa Maria, 110v Valparaiso, Chile,
$^{ll}$Universite catholique de Louvain, 1348 Louvain-La-Neuve, Belgium,
$^{oo}$University of Z\"{u}rich, 8006 Z\"{u}rich, Switzerland,
$^{pp}$Massachusetts General Hospital and Harvard Medical School, Boston, MA 02114 USA,
$^{qq}$Hampton University, Hampton, VA 23668, USA,
$^{rr}$Los Alamos National Laboratory, Los Alamos, NM 87544, USA,
$^{ss}$Royal Holloway, University of London, Egham, Surrey, TW20 0EX, UK
}}
\noaffiliation

\date{\today}

\begin{abstract}
We report on a measurement of the top-quark electric charge in $t\bar{t}$ events in which one $W$ boson originating from the top-quark pair decays into leptons and the other into hadrons. The event sample was collected by the CDF II detector in $\sqrt{s}=1.96$ TeV  proton-antiproton collisions and corresponds to 5.6 fb$^{-1}$. We find the data to be consistent with the standard model and exclude the existence of an exotic quark with $-$4/3 electric charge and mass of the conventional top quark at the 99\% confidence level.
\end{abstract}


\maketitle

\newpage
\section{\label{sec:Intro}Introduction}

Since the discovery of the top quark ($t$) \cite{CDFt,D0t}, the CDF and D0 collaborations, joined recently by the LHC experiments, have measured several of its properties to be consistent with standard model (SM) predictions. Determining that the top quark decays into a $W^+$
boson and a bottom quark ($b$), while the anti-top quark decays to a $W^-$ boson and an
anti-bottom quark would ensure indirectly that the electric charge of the (anti-)top
quark is indeed ($-$)$2/3$ as expected in the SM. 
If events were found to contain decays into a $W^-$ and bottom-quark final state, the charge of the decaying particle would be --4/3, incompatible with the SM top quark. Motivation for a measurement was proposed in Ref.\ \cite{chang}, where such a hypothesis was put forward.  In this model, an exotic quark of mass around $170$ GeV/$c^2$ is assumed to be part of a fourth generation of quarks and leptons, while the standard-model top quark is heavier than $230$ GeV/$c^2$.  Even though this model is by now strongly disfavored by other measurements \cite{CDFstop,D0stop}, the charge correlations between jets initiated by $b$ or $\bar b$ quarks and $W$ bosons in $t\bar t$ events have not yet been definitively established. The existence of an exotic decay combination ($b$ coupled to $W^{-}$ and $\bar b$ coupled to $W^{+}$) has already been constrained experimentally \cite{D0, SLT}, but with less sensitivity than the present measurement.
\newline
\indent
In this article we analyze $t\bar t$ candidate events and treat the SM and exotic-quark hypotheses exclusively. We analyze $t\bar{t}$ candidate events in the final state containing hadrons from the decay of one $W$ boson and an electron or muon and corresponding antineutrino from the decay of the other $W$ boson. We first determine the charge of the $W$ boson (using the charge of the lepton or the opposite charge for the hadronically decayed $W$ boson). Then we pair the $W$ boson with the jet originating from a $b$ quark ($b$ jet) from the same top-quark decay. Finally we determine the charge of the $b$ jet using an optimized jet-charge algorithm, JetQ \cite{Feynman, Barate, Bednar, Unalan}. Pairings where the  charge of the $W$ boson is opposite to the JetQ value are classified as standard-model-like (SM-like) decays, while pairings where
the charge of the $W$ boson is of the same sign are classified as  exotic-model-like (XM-like) decays.
\newline
\indent
In Sec.\ \ref{sec:CDF} we briefly describe the CDF II detector. The data sample and event selection are presented in Sec.\ 
\ref{sec:Data}, and Monte Carlo simulations in Sec.\ \ref{sec:MC}. Section \ref{sec:pairing} discusses the method to pair the $W$ boson with the correct $b$ jet. The JetQ algorithm used to 
assign a charge to the $b$ jet, as well as its calibration using data, are described in Sec.\ \ref{sec:jetq}. The backgrounds and the possible biases they may induce in the measurement are investigated in Sec.\ \ref{sec:Bak}. In Sec.\ \ref{sec:sys} the systematic uncertainties are presented 
while Sec.\ \ref{sec:estimates} explains how the pairing purity and JetQ purity are combined to obtain the signal purity, i.e., the probability of correctly identifying a signal event as coming from the SM or the XM. The statistical treatment of the data is described in Sec.\ 
\ref{sec:stat} and the results are presented and discussed in Sec.\ \ref{sec:results}.

\section{\label{sec:CDF} The CDF II Detector}

The CDF II detector is described in detail in Refs.\ \cite{CDF, CDFII}. The subdetectors most relevant to this measurement are briefly described in this section. The
detector is approximately hermetic over the full angular coverage and is composed of a charged particle tracker embedded in an axial magnetic field of 1.4 T, 
surrounded by electromagnetic and hadronic calorimeters and muon detectors. A cylindrical coordinate system with $z$-axis directed along the proton beam is used. The polar angle $\theta$ is defined with respect to the proton beam direction and $\phi$ is the azimuthal angle about the $z$-axis. 
Pseudorapidity is defined as $\eta = -\ln \tan(\theta/2)$.

The charged particle tracker is composed of silicon micro-strip detectors \cite{L00, SVX, ISL} covering the pseudorapidity range of $|\eta|<2$ and providing 11 $\mu$m spatial resolution on each measurement point in the $r$--$\phi$ plane, crucial for the identification of secondary vertices characteristic of jets originating from $b$ quarks. The silicon detectors are surrounded by a $3.1\,$m long open-cell drift chamber \cite{rCOT}, which measures the momenta of charged particles within a pseudorapidity range of $|\eta|<1$. The calorimeter covers the pseudorapidity range of $|\eta|<3.6$ and is segmented into projective towers 
that point towards the nominal center of the interaction region. The electromagnetic portion is a lead-scintillator sampling calorimeter \cite{cem}, which also contains proportional 
chambers and resistive strips at a depth corresponding to the typical maximum shower intensity for electrons. The hadronic portion is an iron-scintillator sampling calorimeter \cite{cha}. 
Muon detectors are located outside the calorimeters.  Two sets of drift chambers separated by steel absorber, the CMU \cite{cmu} and CMP \cite{cmpcmxI, cmpcmxII}, cover the pseudorapidity range $|\eta|<0.6$, and layers of drift tubes sandwiched between scintillation counters, the CMX \cite{cmpcmxI, cmpcmxII}, cover the range $0.6<|\eta|<1.0$.

\section{\label{sec:Data}Data Sample and Event Selection}

 This analysis is based on a data sample corresponding to an integrated luminosity of 5.6 fb$^{-1}$ collected
with the CDF II detector between February 2002 and February 2010.  The
events first have to pass an inclusive-lepton online event selection (trigger) that requires
 an electron with $E_T>$18 GeV or muon with $p_T>$18 GeV/$c$ \cite{variables}. We then select events offline
with a reconstructed isolated electron $E_T$ (or muon $p_T$) greater
than 20 GeV (GeV/$c$), and missing $E_T  (  {\not\!\!E_T } ) >$20 GeV \cite{MET}.
In addition we require events to have at least four jets, three of them with $E_T>$20 GeV and $|\eta|<2.0$ and another jet with $E_T>12$ GeV and $|\eta|<2.4$. We explicitly reject events that have two or more leptons to ensure that the final sample does not include events where both $W$ bosons decay into leptons (dilepton channel).
\newline
\indent
The electron selection relies on the accurate geometrical match between a reconstructed track and some energy
deposition in the electromagnetic calorimeter. We also require that the amount of energy deposited
in the hadronic calorimeter be significantly less than in the electromagnetic calorimeter. 
An isolation criterion requires the transverse energy in the towers not assigned to the electron, within a cone of $\Delta R \equiv \sqrt{(\Delta{\eta})^2 + (\Delta{\phi})^2} = 0.4$ centered around the lepton, to be less than 10\% of the candidate electron $E_T$.
\newline
\indent
In the muon selection, a track candidate from the tracker is matched to a track segment (stub) in one or more of the muon drift chambers. We require either a stub in both the CMU and CMP chambers, or a stub in the CMX chamber, and refer to the resulting muon candidates 
as CMUP or CMX muons respectively. The energy deposited in the region of the calorimeter to which the trajectory of the candidate muon extrapolates is required to be consistent with the expectation for a minimum-ionizing particle. The isolation criterion for muons, similar to that for electrons, is that the calorimeter transverse energy in a cone of $\Delta R = 0.4$ around the extrapolated muon track (not including 
the muon energy deposition itself) must be less than 10\% of the muon $p_T$. Details on the electron and muon identification are discussed in Ref. \cite{PRD71}.
\newline
\indent
The muon acceptance is increased by approximately 20\% by including events containing muons that cannot be triggered on directly. Such events must pass a different trigger, which requires a missing transverse energy larger than $35\,$GeV and at least two jets of $E_T > 10\,$GeV. Candidates are selected if they contain a CMX stub in a region not covered by the inclusive lepton trigger, or a stub only in the CMU or CMP chambers, or an isolated track not fiducial to any muon detector. Muons in these categories, called \emph{extended muons}, are also required to pass the isolation criterion and to have  $p_T > 20\,$GeV/$c$. Dilepton veto and jet requirements are the same as those applied to events selected from the inclusive lepton trigger. To ensure full efficiency of the trigger, the extended muon candidates are also required to have two jets with 
$E_T > 25\,$GeV, one of which should be central ($|\eta|<0.9$) and separated from the other by $\Delta R>1.0$.
\newline
\indent
The jet reconstruction is based on a calorimeter-tower-clustering cone algorithm with a cone size of $\Delta R=0.4$. Towers corresponding to selected electrons are removed before clustering. The observed $E_T$ for jets is corrected for the effects of jet fragmentation, calorimeter non-uniformities and the calorimeter absolute energy scale \cite{JES}. 
\newline
\indent
Due to the presence of a neutrino leaving the detector undetected, there will be an imbalance in the 
transverse momentum of the event. Consequently, events are expected to have some missing transverse energy ${\not\!\!E_T}$, and we require ${\not\!\!E_T} > 20$ GeV.
\newline
\indent
The data set selected above, called ``lepton+jets" (LJ), is dominated by
QCD production of $W$ bosons with multiple jets (``$W$ + jets").  To improve the
signal-to-background ratio we identify events with two or more $b$ jets, i.e., we
require at least two of the jets to contain a secondary vertex, characteristic of
a $B$ hadron having decayed. This secondary vertex algorithm is tuned such that the efficiency of identifying a $b$ jet is about 50\%, 
and results in a probability of about 2\% of misidentifying a light-quark jet. More information about this algorithm can be found in Ref.\ \cite{PRD71}. 

\section{\label{sec:MC} Monte Carlo Simulation}

The $t\bar{t}$ Monte Carlo (MC) simulation used in this measurement relies on {\sc pythia} version 6.216 \cite{pythia} for event generation
and parton showering. The top-quark mass used is $172.5\,$ GeV$/c^2$. Samples generated with other values
of the top-quark mass are studied for any dependence of the measurement on this parameter. A sample of $t\bar{t}$ events generated with {\sc herwig} version 6.510 \cite{herwig} is used to estimate a possible systematic uncertainty due to the choice of generators. Most of the background samples rely on {\sc pythia} except for the $W$ + jets background, which is generated using transition matrix elements calculated by {\sc alpgen} version 2.10$^{\prime}$ \cite{alpgen} and {\sc pythia} for parton showering. Parton distribution
functions are modeled with CTEQ5L \cite{cteq5l}. The interactions of particles with the detectors are modeled using {\sc geant3} \cite{geant}, and the {\sc gflash} parametrization \cite{gflash} for showers in the calorimeters. Details on the implementation and tuning of the CDF II detector simulation are found in Ref.\ \cite{simCDF}. 

\section{\label{sec:pairing} Pairing between the $W$ boson and the $b$ jet}

Each event contains a lepton, multiple $b$-jet candidates, and non-$b$ jets. In order to assign the four highest-$p_T$ jets to the four final-state quarks from the $t\bar{t}$ decay and to associate the lepton with the $b$ jet from the decay of the top quark that produced the leptonically-decaying $W$ boson, we use the top-quark mass kinematic fitter described in Ref.\ \cite{topmass}, which minimizes a $\chi^2$ variable that incorporates constraints on the top-quark mass $m_t$, fixed at $172.5\,$GeV/$c^2$, and on the $W$-boson mass $m_W$, fixed at $80.42\,$GeV/$c^2$. The $\chi^2$ is given by

\begin{widetext}
\begin{eqnarray}
\chi^2 = \sum\limits_{\mathit{i=\ell,4jets}}\frac{ (\hat{p}_T^i - p_T^i )^2  }{\sigma_i^2} +
 \sum\limits_{\mathit{j=x,y}} \frac{( \hat{p}_j^{\mathit{UE}} - p_j^{\mathit{UE}})^2}{\sigma_j^2} + \frac{(m_{\mathit{jj}} - m_W)^2}{\sigma_W^2} + \frac{( m_{\mathit{\ell \nu}} - m_W)^2}{\sigma_W^2}
+ \frac{(m_{\mathit{bjj}} - m_t)^2}{\sigma_t^2} + \frac{( m_{\mathit{b\ell \nu}} - m_t)^2}{\sigma_t^2} \cdot
\end{eqnarray}
\end{widetext}

\noindent
The first term evaluates the difference between the best-fit value ($\hat{p}_T$) and the observed value ($p_T$) of the transverse
momentum for the four highest-$p_T$ jets and the lepton. The second term evaluates the difference between the best-fit and the observed
value of the unclustered energy, which represents the energy in the calorimeter towers not associated with the jets or primary lepton. The last four terms represent the mass differences between the $W$ boson and its decay products and between the top quark and its decay products. The parameter $m_{t}$ is not floating, in contrast to Ref.\ \cite{topmass}. The $\sigma_{i}$ and $\sigma_{j}$ variables are the uncertainties on the observed momenta values, while $\sigma_{W}$ represents the decay width of the $W$ boson (2.12 GeV/$c^{2}$), and $\sigma_{t}$ is the quadrature sum of the theoretical width of the top quark (1.5 GeV/$c^{2}$) and the experimental uncertainty on its mass (0.9 GeV/$c^{2}$). Since events may contain two, three, or four jets identified as $b$ jets by the secondary-vertex algorithm, 
there are two, six, or twelve possible assignments of $b$ jets to $W$ bosons, respectively. For each $W-b$ pairing two $\chi^2$ values are computed to allow for the unknown $z$ component of the neutrino momentum. Choosing the combination that minimizes this $\chi^2$ leads to a purity $p_{\mathit{pair}}$ (the probability of correct $W-b$ pairing) of 76\%, as estimated with the {\sc pythia} $t\bar{t}$ MC sample. By imposing an upper threshold to the value of the minimum $\chi^2$, the purity is increased but the event selection efficiency is reduced. We identify the optimal configuration by maximizing $\epsilon D^{2}$ obtained from the $t\bar{t}$ simulated sample, where $\epsilon$ is the efficiency of the $\chi^{2}$ requirement and $D$ is the dilution, defined as $D\equiv 2p_{\mathit{pair}}-1$. By restricting the analysis to events in which the minimum $\chi^2$ does not exceed 9, we achieve an efficiency on signal of $53.2\pm 0.1$\% with a purity $p_{\mathit{pair}}$ of $83.3\pm 0.1$\%.

\section{\label{sec:jetq} Charge of a  $b$ jet}

We use the jet-charge (JetQ) algorithm to determine which of the high-$p_T$ $b$ jets characteristic of a $t\bar{t}$ event originated from a $b$ quark, and which from a $\bar{b}$ quark.  We select tracks with impact parameter \cite{impactParam} less than 0.15 cm with respect to the primary vertex and  $p_T$ larger than 1.5 GeV/$c$ within a cone of $\Delta R <0.4$  around the $b$ jet axis. We only compute JetQ if there are at least two such tracks within this cone. We then sum up the charges of those tracks with weights that depend on
their momentum component along the jet axis:

\begin{equation}
JetQ = \frac{\sum(\vec{p}_{\mathit{track}} \cdot \vec{p}_{\mathit{jet}})^{0.5}Q_{\mathit{track}}}{\sum(\vec{p}_{\mathit{track}} \cdot \vec{p}_{\mathit{jet}})^{0.5}},
\end{equation}

\noindent
where $\vec{p}_{\mathit{jet}}$ ($\vec{p}_{\mathit{track}}$)  is the momentum vector of the jet (track) and $Q_{\mathit{track}}$ is the charge of the particle associated to the track. Track requirements and the choice of the 0.5 exponent result from an optimization of JetQ  on the simulated $t\bar{t}$ sample. If the JetQ value is
positive we assign the bottom jet to a $\bar{b}$ quark, if it is negative we assign the bottom jet to a $b$
quark. Monte Carlo studies indicate that this algorithm has a selection efficiency of $97.9\pm 0.1$\% and a purity per identified $b$ jet of about $60.8\pm 0.1$\%.
\begin{figure*}[ht!]
 \begin{center}
    \includegraphics[width=18cm,clip=]{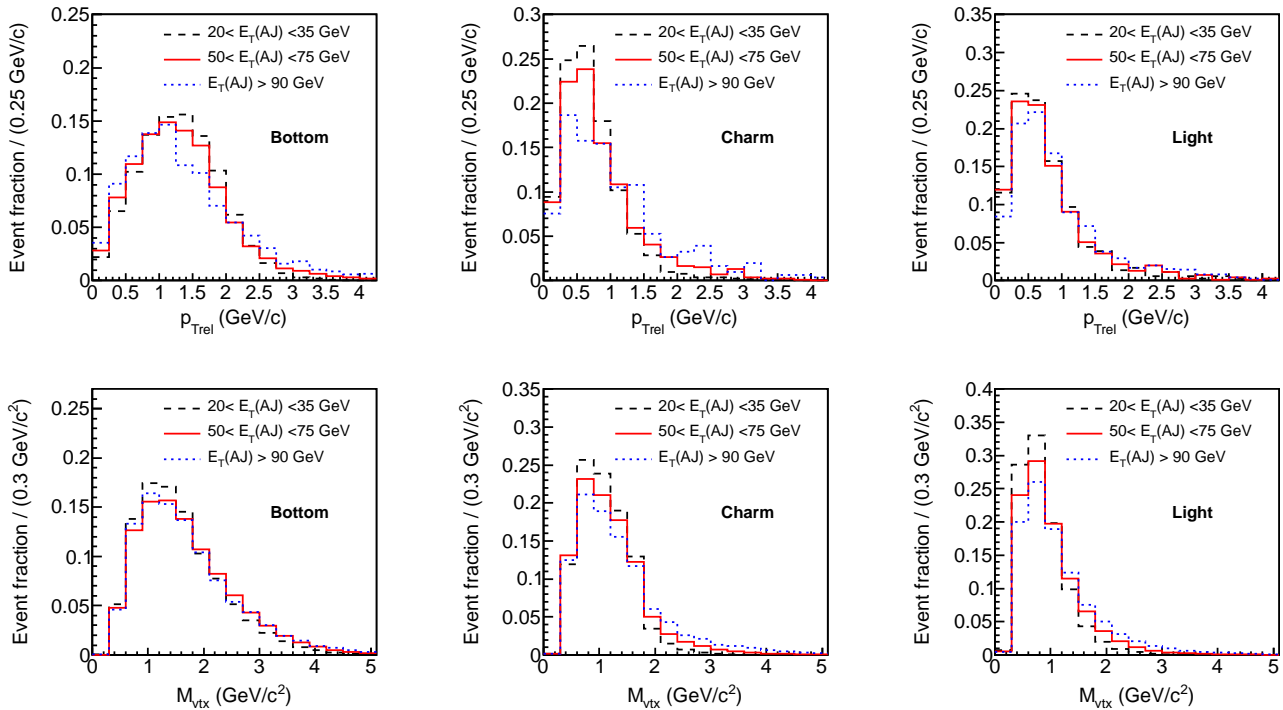}
    \caption{\label{fig:ptrel}Examples of $p_{Trel}$ (top) and secondary-vertex mass (bottom) templates for bottom, charm, and light quarks for different values of away-jet $E_T$.}
  \end{center}
\end{figure*}

\subsection{\label{sec:calib}Calibration of the JetQ purity in data}

Since the simulation does not model the jet fragmentation reliably, we correct the purity of the JetQ algorithm obtained from the simulation by using a dijet data sample enriched in heavy flavor.
This sample is collected with a trigger that requires a central muon with $p_{T} >8$ GeV/$c$. Events are required to have a tag \emph{muon jet} with $E_T > 20$ GeV that contains a muon  with $p_T >9$ GeV/$c$ inside the cone, and a probe \emph{away jet} with  $E_T > 20$ GeV and with $\Delta\phi > 2$ with respect to the muon jet. We require both jets to be identified as $b$ jets using the secondary-vertex algorithm, but a more selective variant of the tagger is used for the muon jet. The JetQ purity is obtained as the fraction of selected events in which the charge of the muon is opposite to the JetQ value of the away jet.
 The observed purity is corrected for a number of effects. If the muon originates from a $b \to c \to \mu$ cascade decay, its charge is the opposite of the one it would have if coming directly from a $b$ decay (secondary fraction); if the $B$ meson undergoes mixing, the charge of the muon may also flip sign (mixing fraction); and finally, if one of the two $b$ jet candidates is misidentified, no correlation between the JetQ value and the charge of the identified muon (non-$b\bar{b}$ fraction) is present. The first two effects can be obtained from simulation. The last effect is calculated from a fit of simulation to data. 
\newline
\indent
In order to obtain the $b\bar{b}$ fraction of the dijet sample, we use two independent fits. We first extract the $b$ fraction in muon jets by fitting the distribution of $p_{\mathit{Trel}}$, the component of the muon momentum transverse to the jet direction, which is enhanced at larger values for muons originating from $b$ quark jets. Figure \ref{fig:ptrel} (upper panels) shows a selection of the $p_{\mathit{Trel}}$ templates used. For this fit we combine the charm and light-quark templates since they are very similar. Then, we determine the $b$ fraction in away jets by fitting the secondary-vertex-mass distribution, which is enhanced at higher values when the parent quark is heavier. Figure \ref{fig:ptrel} (lower panels) shows a selection
of secondary-vertex mass templates used; the template shapes depend on the away-jet $E_T$. To allow for the possibility that the simulated sample might not model reliably the $E_T$ distribution of light quarks 
\begin{figure}[ht!]
 \begin{center}
   \includegraphics[width=8.5cm,clip=]{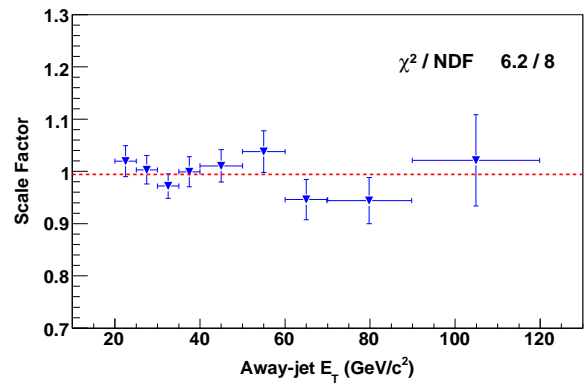}
    \caption{\label{fig:SF}Ratio between the purity of the JetQ algorithm in data and Monte Carlo simulation as a function of away-jet $E_T$. The dotted line is the result of fitting a constant through the data points.}
  \end{center}
\end{figure}
that are misidentified as $b$ quarks, we perform all template fits in nine independent ranges of away-jet $E_T$. Since the $b$ fractions of the muon and away jets are obtained from independent fits, we have no  information on their correlation in the dijet sample.  However we can obtain the highest (lowest) value of the $b\bar{b}$ fraction by assuming that this correlation is maximal (minimal). We then estimate the $b\bar{b}$ fraction in each $E_{T}$ range as the average of the upper and lower limits in the range, and set the corresponding uncertainty to half the difference between the limits.

\begin{table*}[htbp]
  \begin{center}
  	\caption{\label{tab:backgpred}Background and signal expectations before and after the $\chi^2$ and JetQ criteria (columns 2 and 5). The efficiencies of these criteria are shown in columns 3 and 4.
The last column includes a factor of two because each selected event contains two $W-b$ pairs, each providing a ``quark candidate" (SM- or XM-like candidate). The uncertainties are discussed in Sec. \ref{sec:sys}.}
		\begin{tabular}{lcccc}
			\hline \hline
			\multirow{2}{*}{Process} & Events           & $\chi^2$ requirement & JetQ       & Quark candidates \\
						 & before criteria  & efficiency   & efficiency & after criteria          \\

			\hline
			
			$W$+HF & $66  \pm 22$ & $0.150 \pm 0.004$ & $0.970 \pm 0.003$ & $19.5 \pm 6.4$\\
			
			QCD fakes & $18 \pm 14$ & $0.17 \pm 0.08$ & $0.88 \pm 0.12$ & $ 5.4 \pm 4.8$\\
			
			Diboson & $4.7 \pm 0.7$ & $0.22 \pm 0.02$ & $0.97 \pm 0.01$ & $ 2.0 \pm 0.4 $\\
			
			Mistag & $9.7 \pm 2.6$ & $0.15 \pm 0.02$ & $0.96 \pm 0.02$ & $ 2.8 \pm 0.8 $\\
			
			Single top & $10.6 \pm 1.3$ & $0.210 \pm 0.004$ & $0.970 \pm 0.003$ & $4.4 \pm 0.5$\\
			
			\hline
			Total background & $109\pm 26$ & - & - & $34 \pm 8$\\
			
			\hline
			Signal & $670 \pm 110$ & $0.532  ^{\pm 0.001({\rm stat})}_{\pm 0.005({\rm syst})}$ & $0.979  ^{\pm 0.000({\rm stat})}_{\pm 0.002({\rm syst})} $ & $700 \pm 120$\\

			\hline \hline

		\end{tabular} 	
	\end{center}

\end{table*}

\begin{table*}[htbp]
	\begin{center}
		\caption{\label{tab:backgcorr} Correlation between lepton charge and JetQ in background and signal events. The last two columns show the expected numbers of SM-like and XM-like quark candidates.}
		\begin{tabular}{lcccc}
		\hline \hline

			\multirow{2}{*}{Process} & Expected number           & \multirow{2}{*}{Correlation} & \multirow{2}{*}{SM} & \multirow{2}{*}{XM}\\			
			                         & of quark candidates &             &     &       \\			

			\hline
			$W$+HF & $19.5 \pm 6.4$ & $0.5 \pm 0.0$ & $9.7 \pm 3.2$ & $9.7 \pm 3.2$\\
			
			QCD fakes & $5.4 \pm 4.8$ & $0.48 \pm 0.06$ & $2.6 \pm 2.3$ & $2.8\pm 2.5$\\
			
			Diboson & $2.0 \pm 0.4$ & $0.5 \pm 0.0$ & $1.0 \pm 0.2$ & $1.0 \pm 0.2$\\
			
			Mistag & $2.8 \pm 0.8$ & $0.5 \pm 0.0$ & $1.4 \pm 0.4$ & $1.4 \pm 0.4$\\
			
			Single top & $4.4 \pm0.5$ & $0.51 \pm 0.01$ & $2.3 \pm 0.3$ & $2.2 \pm 0.3$\\
			
			\hline
			Total background & $34 \pm 8$ & $0.50\pm 0.01$ & $17 \pm 4$ & $17 \pm 4$\\
	                   
			\hline
			Signal & $700 \pm 120$ & $0.562^{\pm 0.004({\rm stat})}_{\pm 0.011({\rm syst})}$ & $394 \pm 66$ & $306 \pm 51$\\
			\hline \hline
			
		\end{tabular}
	
	\end{center}

\end{table*}	

Combining the $b\bar{b}$ fraction with the secondary and mixing fractions we correct the bias in the measured purity in each away-jet $E_T$ bin. We compute a scale factor $SF_{\mathit{JQ}}$ as the ratio of the purity obtained in the dijet data sample to that obtained in a corresponding simulated sample. We see no dependence of the scale factor on the away-jet $E_T$, as shown in Fig. \ref{fig:SF}. We estimate a total systematic uncertainty on the JetQ scale factor of $3.2\%$, coming from uncertainties on the template shape ($2.3\%$), the fit strategy ($1.8\%$), and the $E_T$ dependence ($1.4\%$). We obtain a value of the scale factor of $SF_{\mathit{JQ}}=0.99\pm0.01 ({\rm stat}) \pm 0.03 ({\rm syst})$. 

\section{\label{sec:Bak}Backgrounds}

\begin{figure}[ht!]
 \begin{center}
    \includegraphics[width=8.5cm,clip=]{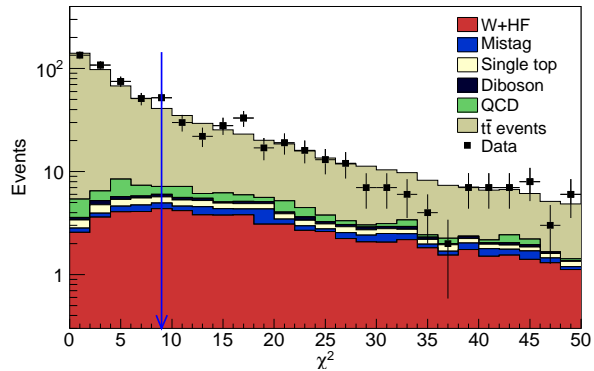}
    \caption{\label{fig:chi2} Distribution of minimum $\chi^2$  for events passing selection requirements described in Sec. \ref{sec:Data}. Shaded histograms show signal and background predictions stacked to form the total prediction. The arrow shows the $\chi^2$ upper threshold.}
  \end{center}
\end{figure}

\begin{figure}[ht!]
 \begin{center}
    \includegraphics[width=8.5cm,clip=]{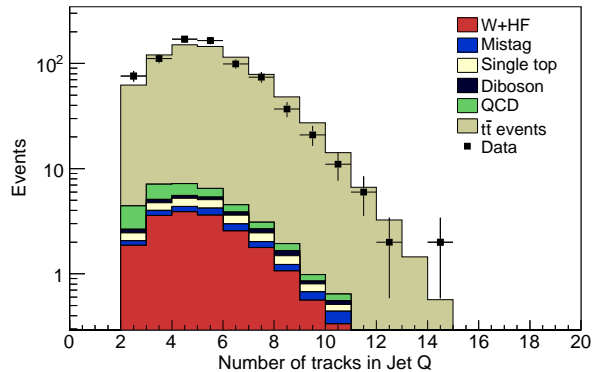}
    \caption{\label{fig:numTrks} Distribution of number of tracks entering the JetQ calculation. Shaded histograms show signal and background predictions stacked to form the total prediction. The purity of the JetQ algorithm is calibrated as described in Sec. \ref{sec:calib} and a scale factor to the Monte Carlo is obtained to account for modeling discrepancies.}
  \end{center}
\end{figure}

\begin{figure}[ht!]
 \begin{center}
    \includegraphics[height=6cm,width=8.5cm,clip=]{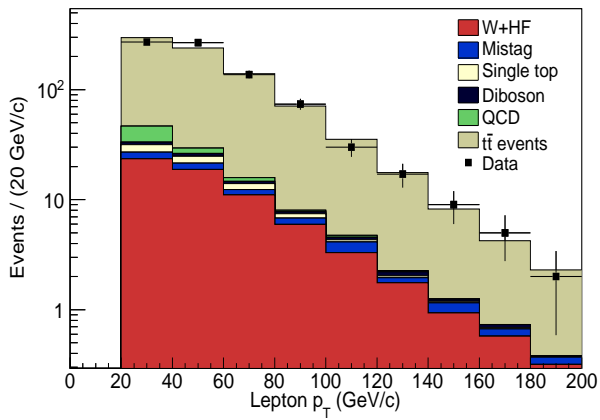}
    \caption{\label{fig:leppt} Lepton $p_T$ distribution. Shaded histograms show signal and background predictions stacked to form the total prediction.}
  \end{center}
\end{figure}

In the following, \emph{signal} refers to events with either a SM $t\bar{t}$ pair, or a pair of exotic quarks with mass $172.5\,$GeV$/c^2$. The exotic quarks are simulated using the standard $t\bar{t}$ Monte Carlo described in Sec.\ \ref{sec:MC}. The dominant background is QCD production of $W$ plus multijet events.  These events enter the signal sample when two of the jets  are $b$ jets ($W$+HF), or light quark
jets are misidentified as $b$ jets (mistag). Other backgrounds include QCD multijet events where a jet is misidentified as a lepton and two jets are $b$ jets or misidentified as such (QCD fakes), single-top-quark events, and diboson events. The amount of background is moderate ($\approx 15$\%) because at least two jets are required to be identified as $b$ jets.
\newline
\indent
We obtain the background predictions with the same method as for the cross-section measurement of Ref.\ \cite{LJcross}. We compute the efficiency of the $\chi^2$ requirement and JetQ selection using Monte Carlo simulation for each background with the exception of the QCD fakes, for which we use data. Finally, we search for correlations between the charge of the primary lepton and the JetQ value of the corresponding $b$ jet in each background source. This correlation is expressed as the fraction of the total number of $W$--$b$ pairs that are classified as SM-like. We expect this fraction to be 50\%, i.e., the same probability for pairs to be SM- or XM-like, except for two processes, single-top-quark production and QCD $b\bar{b}$ production where a lepton from the semileptonic $b$ decay is misclassified as primary lepton. For the first process we rely on the simulation to estimate the possible correlation, while for the second process we use a data sample where all the LJ selection requirements are applied except those of the lepton selection, and we require instead that the lepton fail at least two identification criteria. The composition of this sample is dominated by QCD background events. Table \ref{tab:backgpred} summarizes the signal and background predictions.
Table \ref{tab:backgcorr} summarizes the amount of correlation for each background. Background sources for which no effect is expected are assigned a correlation of 0.5. The signal correlation (purity) is defined in Sec.\ \ref{sec:estimates}.
\newline
\indent
In Fig. \ref{fig:chi2} we show the $\chi^2$ distribution used to assign the lepton to the correct $b$ jet, while in Figs. \ref{fig:numTrks} and \ref{fig:leppt} we show the distributions 
of the number of tracks in the JetQ calculation and the lepton $p_T$, respectively. 

\section{\label{sec:sys}Systematic Uncertainties}

Systematic uncertainties come from modeling of the geometrical and kinematic acceptance, knowledge of
the secondary vertex tagging efficiency, the effect on the acceptance of the uncertainty on the jet energy scale, uncertainties on the background predictions, and the uncertainty on the luminosity.
\newline
\indent
Monte Carlo modeling of geometrical and kinematic acceptance includes effects of parton distribution functions (PDFs), initial- and final-state radiation, and jet energy scale. The PDF uncertainty is estimated by varying the independent eigenvectors of the CTEQ6M \cite{cteq6} PDF set, varying the QCD scale ($\Lambda_{QCD}$), and comparing the nominal CTEQ5L \cite{cteq5l} PDF set with MRST72 \cite{mrst725}. We vary the parameters that govern initial- and final- state radiation and obtain the corresponding uncertainty by comparing the results with the nominal one. Similarly the uncertainty coming from jet energy scale is estimated by varying the scale within its uncertainties. An additional systematic source comes from the choice of the generator (and in particular the hadronization model), for which we compare {\sc pythia} with {\sc herwig}.
\newline
\indent
All of these systematic uncertainties affect the predicted numbers of
signal and background events (for details see Ref.\ \cite{LJcross}) and the efficiency and purity of the pairing and JetQ algorithms. An additional systematic uncertainty affects the pairing: the choice of the top-quark mass used in the simulated sample and in the $\chi^2$ constraint. We measure this uncertainty from the shift of the values obtained when comparing the nominal results to those from two extra samples generated with top-quark masses of 170 and 175 GeV/$c^2$. Finally, for the JetQ purity systematic uncertainty,
we take the value obtained from the calibration in data and add in quadrature the effect of initial- and final-state radiation, since these may be different between a $b\bar{b}$ and a $t\bar{t}$ environment.
In Table \ref{tab:syst} we show the systematic uncertainties on the pairing efficiency and purity, and on the JetQ selection efficiency and purity. These systematic uncertainties are assigned only to the signal as for backgrounds the statistical uncertainty is dominant.

\begin{table*}
	\begin{center}
		\caption{\label{tab:syst} Summary of systematic uncertainties (in \%) for the $\chi^{2}$ selection and JetQ efficiencies, and
    		for the pairing and JetQ purities. The ($0.7$) figure is given 
	as information but not used in the total uncertainty since the JetQ purity is calibrated in data and the 
	corresponding scale factor already corrects for the Monte Carlo hadronization model. The total uncertainty is calculated as a sum in quadrature of the individual uncertainties coming from the different sources.}
		\begin{tabular}{ccccc}
		\hline \hline
		Systematic (in $\%$) &  $\chi^2$ selection efficiency & JetQ efficiency & Pairing purity & JetQ purity\\
		\hline
		Jet energy scale& $0.2$ & $0.04$ & $0.1$ & $0.1$\\
		Initial- and final-state radiation & $0.5$ & $0.1$ & $0.2$ & $0.2$\\
		MC generator & $0.2$ & $0.1$ & $0.1$ & ($0.7$)\\
                Top-quark mass & $0.4$ & $0.2$ & $0.9$ & $0.5$\\
		PDF & $0.7$ & $0.02$ & $0.1$ & $0.02$\\

		\hline
		Total & $1.0$ & $0.3$ & $1.0$ & $0.6$\\
		\hline \hline
		\end{tabular}
	
	\end{center}

\end{table*}

\section{\label{sec:estimates} Signal Purity Determination}
In Table \ref{tab:backgcorr} we show the signal purity that leads to the estimation of the expected numbers of SM-like and XM-like quark candidates. The purity is a combination of the pairing purity and the JetQ purity as follows:

\begin{widetext}
\begin{eqnarray}
p_s=f_{\mathit{nb}}\,SF_{\mathit{nb}}\,p_{\mathit{nb}}+(1-f_{\mathit{nb}}\,SF_{\mathit{nb}})[p_{\mathit{Wb}}\,p_{\mathit{JQ}}\,SF_{\mathit{JQ}}+(1-p_{\mathit{Wb}})(1-p_{\mathit{JQ}}\,SF_{\mathit{JQ}})],
\label{eq:sigPur}
\end{eqnarray}
\end{widetext}

\noindent
where $f_{\mathit{nb}}$ is the fraction of signal Monte Carlo events where we have
misidentified the $b$ jet and $SF_{\mathit{nb}}$ is a scale factor that accounts for any difference in the rate of misidentified $b$ jets between data and simulation. This is the same scale factor determined
in the measurement of the top-quark-pair production cross-section using $b$-jet 
tagging \cite{LJcross}. The quantity $p_{\mathit{nb}}$ is the probability that a signal event with a misidentified $b$ jet will be correctly labeled as SM- or XM-like, $p_{\mathit{Wb}}$ is the pairing purity for
cases where the JetQ was defined, and $p_{\mathit{JQ}}$ is the JetQ purity for
the cases where the pairing criterion was applied. These three purities are obtained from simulated events.  The $SF_{\mathit{JQ}}$ is the
scale factor between data and Monte Carlo simulation for the JetQ obtained from the data
calibration study (see Sec. \ref{sec:calib}). Table \ref{tab:combpur} shows the values used in Eq.\ (\ref{eq:sigPur}), with uncertainties propagated from those in Table \ref{tab:syst}.
\begin{table}[ht!]
 \begin{center}
 	\caption{\label{tab:combpur}Inputs to the signal purity.}
	\begin{tabular}{lc}
	 \hline \hline
	 $f_{nb}$ & $0.079 \pm 0.001$ \\
	 $SF_{nb}$ & $1.01 \pm 0.03$ \\
	 $p_{nb}$ & $0.50 \pm 0.01$ \\
	 $p_{Wb}$ & $0.833 \pm 0.001({\rm stat}) \pm 0.008({\rm syst})$  \\
	 $p_{JQ}$ & $0.608 \pm 0.001({\rm stat}) \pm 0.003({\rm syst})$  \\
	 $SF_{JQ}$ & $0.99 \pm 0.01({\rm stat}) \pm 0.03({\rm syst})$ \\
	 \hline \hline
	 \end{tabular}
	
	\end{center}

\end{table}	

The equivalent of signal purity for  background events is the correlation between JetQ and the primary-lepton charge, and is provided in Table \ref{tab:backgcorr}. Finally, Table \ref{tab:final} summarizes the important analysis inputs to the statistical extraction of results described in the next section.
\begin{table}[h!]
	\begin{center}
	\caption{\label{tab:final}Estimated numbers of background and signal candidates together with the corresponding  purities.}

		\begin{tabular}{cc}
		\hline \hline
		$N_{s}$ & $700 \pm 120$\\
		$N_{b}$ & $34 \pm 8$\\
		$p_{s}$ & $0.562 \pm 0.004 ({\rm stat}) \pm 0.011 ({\rm syst})$\\
		$p_{b}$ & $0.50 \pm 0.01$\\
		\hline \hline
		\end{tabular}
	
	\end{center}
\end{table}	

\section{\label{sec:stat} Statistical treatment}
Once we apply the pairing and JetQ selection to the data, we classify each data pair as SM-like or XM-like, and define $f_{+}$ to be the  fraction of SM candidates among the data pairs. The aim of the measurement is to test the SM hypothesis ($f_{+}=1$) against the XM hypothesis ($f_{+}=0$). We write the likelihood as the product of two Poisson probabilities for the observed numbers $x^{+}$ and $x^{-}$ of SM- and XM-like candidates (respectively), and four Gaussian constraints on the nuisance parameters $y_{s}$, $y_{b}$, $z_{p_{s}}$, and $z_{p_{b}}$ (the numbers of signal and background candidates and the purities of signal and background, respectively):

\begin{widetext}
\large{
\begin{eqnarray}
L = \frac{(N_+)^{x^+}e^{-N_+}}{x^+!}\,\frac{(N_-)^{x^-}e^{-N_-}}{x^-!} \, \frac{e^{-\frac{(y_b-N_b)^2}{2\sigma_{N_b}^2}}}{\sigma_{N_b}} \,
\frac{e^{-\frac{(y_s-N_s)^2}{2\sigma_{N_s}^2}}}{\sigma_{N_s}} \,
\frac{e^{-\frac{(z_{p_s}-p_s)^2}{2\sigma_{p_s}^2}}}{\sigma_{p_s}}\,
\frac{e^{-\frac{(z_{p_b}-p_b)^2}{2\sigma_{p_b}^2}}}{\sigma_{p_b}},
\end{eqnarray}
}
\end{widetext}

\begin{figure}[ht!]
 \begin{center}
    \includegraphics[width=8.5cm,clip=]{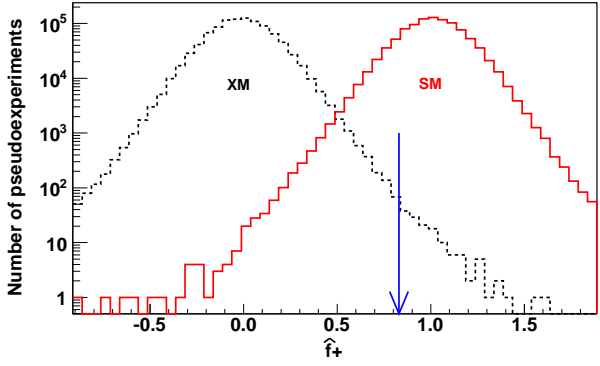}
    \caption{\label{fig:f+}Distribution of the maximum-likelihood estimate of the SM fraction $\hat{f}_+$ from pseudoexperiments
under the XM (dashed line) and the SM (solid line) hypothesis. The arrow shows our result. }	
  \end{center}
\end{figure}

\begin{figure}[h!]
 \begin{center}
     \includegraphics[width=8.5cm,clip=]{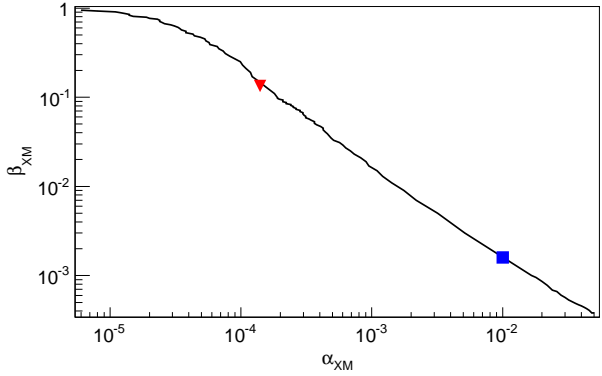}
    \caption{\label{fig:beta} Variation of $\beta_{XM}$ (the probability of accepting a false XM)
      with $\alpha_{XM}$ (the probability of rejecting a true XM). The square represents our {\it a
      priori} choice of $\alpha_{XM}=1\%$, corresponding to $\beta_{XM}=0.16\%$, while the triangle represents the observed $p$-values and is plotted at the coordinates $(p_{XM},p_{SM})$.}
 \end{center}
\end{figure}
\noindent
where $N_{+}$ and $N_{-}$ are the predicted numbers of SM-like and XM-like candidates, and $N_{s}$, $N_{b}$, $p_{s}$, and $p_{b}$ are independent estimates of the nuisance parameters (see Table \ref{tab:final}). The expectations $N_+$ and $N_-$ are computed using the following equations:
\begin{eqnarray}
N_+& = & z_{p_s}y_sf_+ + (1-z_{p_s})y_s(1-f_+) + z_{p_b}y_b \text{ ,}\\
N_-&=& (1-z_{p_s})y_sf_+ + z_{p_s}y_s(1-f_+)+(1-z_{p_b})y_b \text{ .}
\end{eqnarray}

In Fig. \ref{fig:f+} we show the distribution of the maximum-likelihood estimate $\hat{f}_+$  of $f_{+}$, as 
obtained from pseudoexperiments based on either the SM hypothesis or the XM hypothesis. We compute two $p$-values based on $\hat{f}_{+}$ as test statistic: $p_{\mathit{SM}}$ ($p_{\mathit{XM}}$) - the probability of observing a value of the test statistic as in data or smaller (larger) assuming that the SM (XM) hypothesis is true. To reject the SM we require $p_{SM}\le\alpha_{SM}$, where $\alpha_{SM}$ is the standard 5-sigma discovery threshold of $2.87\times 10^{-7}$. To exclude the XM we similarly require $p_{XM}\le\alpha_{XM}$. We note that increasing $\alpha_{XM}$ makes it easier to exclude the exotic model, but reduces the exclusion confidence level $1-\alpha_{XM}$. To optimize the choice of $\alpha_{XM}$ while taking into account the sensitivity of the measurement, we generate pseudoexperiments to compute the probability $\beta_{XM}$ of not excluding the XM when the SM is true, as a function of $\alpha_{XM}$ (Fig.\ \ref{fig:beta}). Using this curve we set $\alpha_{XM}$=1\%, slightly above the value for which $\beta_{XM}$($\alpha_{XM}$)=$\alpha_{XM}$. 

We also quote a measure of evidence based on the data actually observed in the form of a Bayes factor BF,  which is the ratio of posterior to prior odds in favor of the SM. The BF can also be written as the ratio of the likelihood of the SM to the likelihood of the XM. The numerator and denominator are separately integrated over uniform priors for the nuisance parameters. The quantity $2\ln({\rm BF})$ can be interpreted according to a well-established scale \cite{BF}. 

\section{\label{sec:results}Results and Discussion}

In Table \ref{tab:results} we show the number of events and candidates after applying the
pairing and JetQ selection and the number of candidates corresponding
to the SM and XM hypotheses.
\begin{table}[h!]
	\begin{center}
	\caption{\label{tab:results}Observed number of events before and after the pairing requirement, observed number of quark candidates with identified jet charge, and observed  SM-like and XM-like candidates.}
		\setlength{\tabcolsep}{1em}
		\begin{tabular}{cc|ccc}
		\hline \hline
		\multicolumn{2}{c|}{Number of events} & \multicolumn{3}{c}{Quark candidates} \\
		Observed & After pairing & JQ defined & SM & XM \\ \hline
		
		 $815$  & $397$ & $774$ & $416$ & $358$ \\	
		\hline \hline
		\end{tabular}
	
	\end{center}
\end{table}
 
\begin{figure}[h!]
 \begin{center}
    \includegraphics[width=8.5cm,clip=]{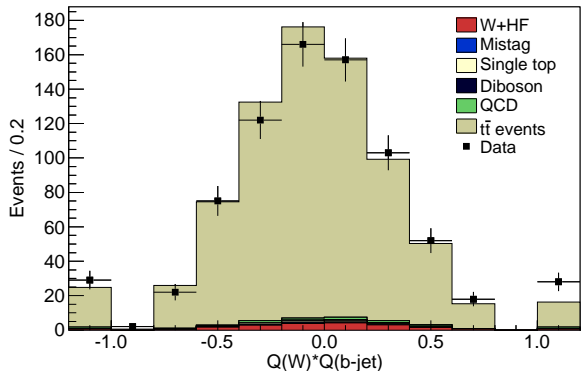}
    \caption{\label{fig:wqjq_both}Distribution of the product of the $W$-boson charge times the JetQ value. Shaded histograms show signal and background predictions stacked for the total prediction. SM-like candidates are on
the negative side of the plot while XM-like candidates are on the positive side. The outermost bins correspond to the cases where JetQ is exactly $\pm1$.}
  \end{center}
\end{figure}
\begin{figure}[ht!]
 \begin{center}
    \includegraphics[width=8.5cm,clip=]{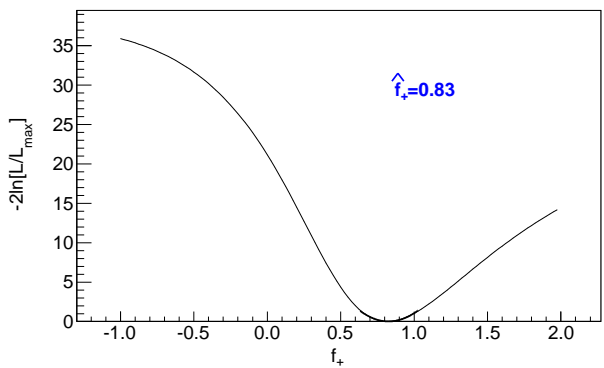}
    \caption{\label{fig:lnl}Distribution of twice the negative logarithm of the profile likelihood as a function of the fraction of SM candidate events in data.}
	
  \end{center}
\end{figure}

\noindent
Candidates whose $W$-boson charge is opposite to the
JetQ value are classified as SM candidates, while candidates whose $W$-boson charge has same sign as the JetQ
are assigned as XM candidates. Figure~\ref{fig:wqjq_both} shows the graphical representation of these numbers, where candidates (and SM expectations) are distributed as function of the product of the JetQ value and the charge of the $W$ boson. Using these numbers we get the profile log-likelihood curve shown in Fig. \ref{fig:lnl}. The minimum of the curve
is at a value of $\hat{f}_{+}=0.83$. This corresponds to a $p$-value of 13.4\% under the SM hypothesis (see red triangle in Fig.~\ref{fig:beta}) and indicates consistency between CDF data and the SM. The $p$-value under the XM hypothesis is 0.014\%, which is interpreted as a 99\% C.L. exclusion of the XM hypothesis. The previous measurements have excluded the XM hypothesis with at most 95\% C.L. \cite{D0, SLT}. We obtain a value of $2\ln(\rm{BF})=19.6$, which, according to the interpretive guidelines of Ref.\ \cite{BF}, constitutes \emph{very strong} evidence in favor of the SM and against the XM.

\begin{figure}[htbp]
 \begin{center}
    \includegraphics[width=8.5cm,clip=]{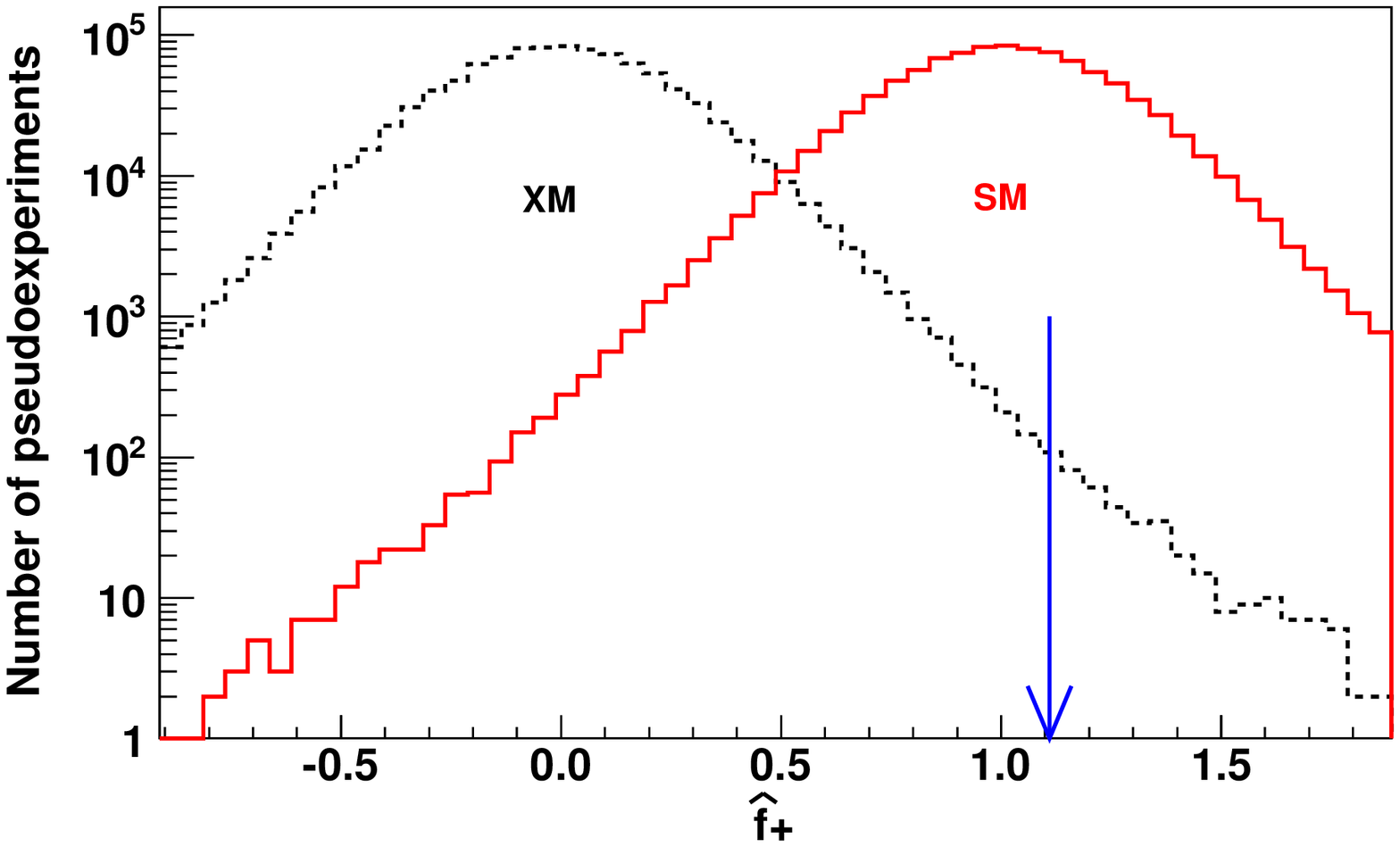}
    \caption{\label{fig:fplusel} Distribution of the maximum-likelihood estimate of the SM fraction $\hat{f}_+$ from pseudoexperiments under the XM (dashed line) and the SM (solid line) hypothesis for electrons only. The arrow shows our result. }
	
  \end{center}
\end{figure}

\begin{figure}[htbp]
 \begin{center}
    \includegraphics[width=8.5cm,clip=]{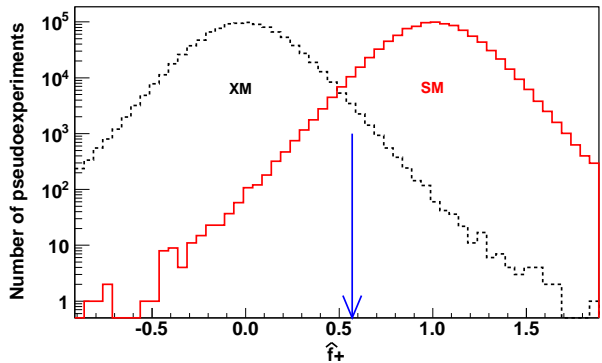}
    \caption{\label{fig:fplusmu} Distribution of the maximum-likelihood estimate of the SM fraction $\hat{f}_+$ from pseudoexperiments
under the XM (dashed line) and the SM (solid line) hypothesis for muons only. The arrow shows our result.}
	
  \end{center}
\end{figure}

Table \ref{tab:elresults} lists the analysis results for electrons and muons separately. In Figs. \ref{fig:fplusel} and \ref{fig:fplusmu} we show the distribution of $\hat{f}_+$ for electrons and muons respectively. Due to its dependence on sample size,  the measurement sensitivity is lower in each lepton subsample than in the full data sample, and an appropriate value of $\alpha_{XM}$ is 5\% in this case. The XM hypothesis is excluded at the 95\% C.L. using the electron or muon subsample. 
\begin{table}[h!]
	\begin{center}
	\caption{\label{tab:elresults} Number of observed candidates and results for the electron and muon candidates separately.}
		\begin{tabular}{lcc}
		\hline \hline
	          & Electrons  & Muons \\ \hline
	         Number of candidates: & $206$ SM and $155$ XM & $210$ SM and $203$ XM \\ 
	         $f_+$ &1.11 & 0.57\\ 
	         $p_{SM}$ &  65.9\% & 5.2\% \\ 
	         $p_{XM}$ & 0.04\% & 0.7\% \\ 
	         $N_s$ & $308 \pm 51$  & $392 \pm 67$ \\ 
	         $N_b$ & $17 \pm 5$ &$17 \pm 4$ \\ 
	         $p_s$ & $0.56 \pm 0.01$ &$0.56 \pm 0.01$  \\ 
	         $p_b$ & $0.50 \pm 0.02$& $0.50 \pm 0.01$ \\ 
	         	\hline \hline
		\end{tabular}
	
	\end{center}

\end{table}

For the muon subsample the $p$-value under the SM hypothesis is only 5.2\%, compared with 65.9\% for the electron subsample. A $\chi^{2}$ test of the hypothesis that the ratio of XM to SM candidates is the same in both subsamples yields a $p$-value of about 9\%, consistent with the discrepancy being a statistical fluctuation.

\section{\label{sec:conclusion}Conclusion}

We present a measurement of the top-quark electric charge that relies on the jet-charge algorithm as an estimator of the electric charge of high-$p_T$ $b$ jets. The measurement uses $t\bar{t}$ pairs reconstructed in final states with one $W$ boson decaying hadronically and the other leptonically, from a data set corresponding to $5.6\,$fb$^{-1}$ of $p\bar{p}$ collisions collected by the CDF II detector. It provides the most sensitive results to support or exclude the exotic-quark hypothesis with $-$4/3 electric charge. Our results exclude this hypothesis at 99\% C.L.. As an additional measure of evidence, the Bayes factor obtained, $2\ln(\rm{BF})=19.6$, supports very strongly the SM over the exotic-quark model hypothesis.

\begin{acknowledgments}
We thank the Fermilab staff and the technical staffs of the participating institutions for their vital contributions. This work was supported by the U.S. Department of Energy and National Science Foundation; the Italian Istituto Nazionale di Fisica Nucleare; the Ministry of Education, Culture, Sports, Science and Technology of Japan; the Natural Sciences and Engineering Research Council of Canada; the National Science Council of the Republic of China; the Swiss National Science Foundation; the A.P. Sloan Foundation; the Bundesministerium f\"ur Bildung und Forschung, Germany; the Korean World Class University Program, the National Research Foundation of Korea; the Science and Technology Facilities Council and the Royal Society, UK; the Russian Foundation for Basic Research; the Ministerio de Ciencia e Innovaci\'{o}n, and Programa Consolider-Ingenio 2010, Spain; the Slovak R\&D Agency; the Academy of Finland; and the Australian Research Council (ARC). 
\end{acknowledgments}


\begin{thebibliography}{99}
\bibitem{variables}
Transverse energy and momentum are defined as $E_T = E\sin(\theta)$ and $p_T = p\sin(\theta)$ respectively.
\bibitem{MET}
The missing transverse energy is defined as $\vec{ \not\!\!E_T} = -\sum\limits_i E_T^i \hat{n}_i$, 
	  where $i$ runs over all calorimeter towers with $|\eta|<3.6$ and
	  $\hat{n}_{i}$ is a unit vector perpendicular to the beam axis and
          pointing to the $i$th calorimeter tower.  The scalar
          quantity ${\not\!\!E_T}$ is then defined as ${\not\!\!E_T}=|\vec{{\not\!\!E_T}}|$.
          Since muons deposit very little energy in the calorimeter, the
          transverse momentum of any identified muon is subtracted from
          $\vec{{\not\!\!E_T}}$, whereas its calorimeter energy is added back in.
\bibitem{impactParam}
The impact parameter is defined as the distance in the transverse plane with respect to the beamline at the track's closest approach.
\end{thebibliography}
\end{document}